\documentclass{revtex4}
\usepackage{graphicx,subfigure,epsfig,epstopdf}
\usepackage{amsmath,amssymb}
\input{epsf}
\input{epsfx}
\newcommand{\D}[2]{\frac{\mbox{d} {#1}}{\mbox{d}{#2}}}

\newcommand{\fr}[2]{\frac{{#1}}{{#2}}}

\newcommand{\unit}[1]{\hspace{-1.3pt} {#1}}

\begin{document}

\title{Electrically controlled modulation in a photonic crystal nanocavity}
\author{Dirk Englund$^{1,2}$, Bryan Ellis$^{1}$, Elizabeth Edwards$^{1}$,  Tomas Sarmiento$^{1}$,  James S. Harris$^{1}$, David A. B. Miller$^{1}$ \& Jelena Vu\v{c}kovi\'{c}$^{1}$}
\address{$^{1}$Department of Electrical Engineering, Stanford University, Stanford CA 94305\\ $^{2}$Department of Physics, Harvard University, Cambridge MA 02138}

%\begin{affiliations}
% \item[1. ] Department of Electrical Engineering, Stanford University, Stanford CA 94305
% \item[2. ] Lyman Laboratory, Harvard University, Harvard MA 02138

%
% \item[] \date \today
%\end{affiliations}

%\section{Introduction}
%Optical inter-and intra-chip connects have the potential to eclipse traditional RF-frequency wires where high-bandwidth, low power communication links are critical [Miller IEEE 2008]. Capacitance, resistance, and clock skew increase as Murphy's law dictates the downscaling of wire interconnects, leading to an inevitable bottleneck.  Consequently there is great interest in developing the key components of optical interconnects, including telecommunications-frequency (C-band) sources, modulators, filters, waveguides, and detectors. Here we present the first electrically pumped photonic crystal refractive modulator. The modulator employs a novel grating to couple normally-incident light into a high quality AlAs/AlGaAs photonic crystal (PC) resonator, resulting in a small device footprint. The resulting low voltage and low capacitance design allows GHz-speed modulation of C-band frequency light.

\begin{abstract}
We describe a compact modulator based on a planar photonic crystal nanocavity whose resonance is electrically controlled. A forward bias applied across a p-i-n diode shifts the cavity into and out of resonance with a continuous-wave laser field in a waveguide. The sub-micron size of the nanocavity promises extremely low capacitance, high bandwidth, and efficient on-chip integration in optical interconnects. 
\end{abstract}

\maketitle

%NOTE : add references to waveguide modulators?

%The design is demonstrated in the GaAs/InGaAs material system, but can be easily adapted to the silicon material system. Under forward-bias the system acts as a light emitting diode with emission channeled into the small spectral linewidth of the cavity; the high spectral power density may be of as a directly modulated light source in optical interconnects.  

%- (Maybe add something about benefits of PC over traditional QW modulator, and/or novelty of design)

% -  photonic crystal nanocavities are particularly interesting because their small footprint comes with a very small capacitance (calculate!),  potentially allowing for modulation speeds on the order of 100GHz(check).
\section{Introduction}
A major challenge in the development of future integrated circuits is increasing the bandwidth of interconnects without raising the power consumption. Replacing electrical with optical interconnects could provide a solution at length-scales down to the chip-to-chip and intra-chip level\cite{2009.Miller,2002.IEEE.Meindl.interconnects,2005.IEEE.Lipson.Si_ic}.  Planar photonic crystals (PPCs) represent an attractive medium for optical interconnects because they allow wavelength-scale guiding of light, efficient multiplexing, and on-chip integration\cite{Joannopoulos08,2006.OpEx.Noda.efficient_drop_filter,2007.APL.Baba.prism}. Optical modulation in PPCs has been achieved by optically injected carriers\cite{2007.NPhoton.Noda.dynamicQ,2007.APL.Fushman.switching,2004NakamuraOptExpr,2006.NPhoton.Notomi.trapping_delaying_high_Q} and direct modulation of photonic crystal lasers\cite{2008.LPR.Englund.Laser_review}.  However, electrically controlled modulation is required for practical devices. Such electrical control was demonstrated recently in a Mach-Zehnder interferometer configuration employing the slow light effect in a photonic crystal waveguide to shrink the active region to $\sim 300$ $\mu$m \cite{2009.OpEx.PC_WG_mod}.  Here we describe a design based on a photonic crystal cavity with a volume of only $\sim 0.8 (\lambda/n)^{3}$, where $n=3.4$ is the refractive index of GaAs at the wavelength $\lambda=1.35\mu$m. The optical transmission of the cavity is shifted by current injection, with a measured modulation visibility up to 35\%. The nanocavity modulator has the potential to operate with an active region on the wavelength-scale, enabling a very compact design with sub-fF capacitance and large modulation bandwidth. Because the cavity modulator relies on a resonance shift rather than absorption, it is expected that it can operate at high optical intensity. 

\begin{figure}
\centering{\includegraphics[width=.7\linewidth]{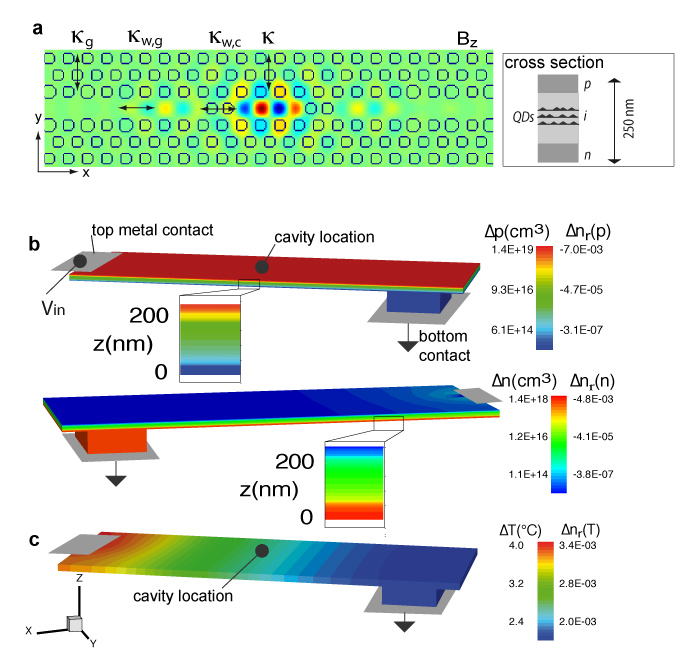}}
\caption{{\footnotesize The photonic crystal circuit. (a) The resonant electromagnetic mode; shown here is the magnetic field component in the $z$-direction, simulated by FDTD.  The cavity energy loss rate is $\kappa$; cavity-waveguide coupling rate $\kappa_{w,c}$; waveguide-grating coupling rate $\kappa_{w,g}$; grating loss rate $\kappa_g$. (b) Steady-state change in hole density $\Delta p$ [top] and electron density  $\Delta n$ [bottom], and associated index change $\Delta n_r(p)[\Delta n_r(n)]$, under $2$V forward bias.  (c) Steady-state change in temperature, $\Delta T$, and associated index change $\Delta n_r(\Delta T)$, under $2$V forward bias. }} \label{fig:setup}
%\vskip2mm
\end{figure}

%Under accumulation conditions  V>0 ,
%the majority carriers in the positive-x silicon region,
%defined in Fig. 1(a), modify the refractive index so
%that phase shift is induced in the optical mode. Compared
%with conventional capacitor-embedded silicon
%rib waveguides [2], the optical mode as indicated in
%Fig. 1(a) is more tightly confined within 1  m perpendicular
%to the capacitor dielectric and thus interacts
%more strongly with the accumulated charges.

\section{Device Design and Fabrication}
The optical device is shown in Fig.\ref{fig:setup}(a). It consists of a modified three-hole defect photonic crystal cavity which is connected on two sides to waveguides in a straight-coupling configuration that is described in previous work\cite{2007.OpEx.Englund}. The cavity has a calculated quality factor of $Q_0=56\cdot 10^{3}$ when it is isolated, estimated by a finite difference time domain (FDTD) simulation.  When the cavity is coupled to the waveguides as shown in Fig.\ref{fig:setup}(a), the simulated quality factor decreases to $Q=9.9\cdot 10^{3}$.  Defining the cavity outcoupling efficiency $\eta$ as the energy decay rate into the waveguides divided by the total decay rate, we estimate $\eta = \kappa_{w,c}/\kappa\approx 0.85$, where $\kappa=\omega/Q,\kappa_0=\omega/Q_0$, and $\kappa_{w,c}=\kappa-\kappa_0$.  Light is coupled out of the plane from the ends of the waveguides by integrated grating structures. %These integrated gratings allow in/out coupling to the waveguide mode with $k_x=\pi/a$, where $a$ is the lattice period\cite{2008.OpEx.Englund}. 

%The system is described by the coupled mode equations (analogous to Ref. \cite{2007.OpEx.Englund}): 
%\begin{equation}
%coupled - mode - equations
%\end{equation}
%Thus we obtain a theoretical coupling efficiency of xxx through the cavity, and coupling efficiency of yy through the full structure (grating-wg-cav-wg-grating). 

%The structure is grown by molecular beam epitaxy and contains three layers of self-assembled InAs quantum dots, which are distributed about the center of the membrane. 

The sample is grown by molecular beam epitaxy (see Appendix \ref{app:growth}). The photonic crystal membrane is illustrated in Fig.\ref{fig:setup}(a) and contains a vertical p-i-n diode for carrier injection. Three quantum dot (QD) layers are used for the characterization of the photonic crystal structure as an internal light source. However, the QDs are not required for the modulation of the signal beam and could be omitted in future designs. The photonic crystal structure was fabricated by a combination of electron beam lithography and wet/dry etching steps, and metal deposition steps for electrodes (Appendix \ref{app:fab}). 
%XXXX ADD SOMETHING LIKE THIS: 
%

\begin{figure}
\centering{\includegraphics[width=\linewidth]{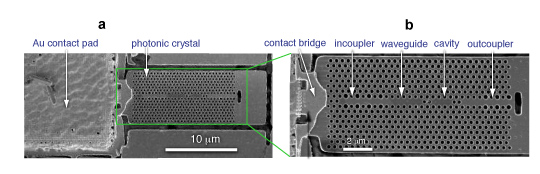}}
\caption{{\footnotesize (a) SEM of the full structure including the contact pad, which connects to the membrane via a bridge across an air trench. (b) Scanning electron micrograph (SEM) of the PPC circuit. The circuit consists of an input grating coupler; input waveguide; cavity; output waveguide; and output grating coupler.  }} \label{fig:setup2}
%\vskip2mm
\end{figure}

Current is injected through a gold bridge to the $p$-layer of the photonic crystal membrane.  The bridge also serves as structural support, holding up one end of the structure (Fig.\ref{fig:setup2}(a,b)).  The other end is supported by an $n$-doped post, which remains after the selective wet etch. The post is connected through the $n$-doped substrate to the sample's bottom metal contact, which is grounded. The signal voltage is applied to the contact pad (Fig.\ref{fig:setup2}(a)). When the structure is forward-biased, electrons flow through the post into the membrane, while holes flow from the bridge into the membrane. We simulated the electrical properties of the structure using the Sentaurus software package, which gives the electron and hole densities, as well as the temperature, as a function of the applied voltage. Fig.\ref{fig:setup}(b) shows the steady-state change in the hole and electron densities under a forward bias of $2.0$\unit{V}.

%. III-V growth - Tomas
%- photonic crystal fab: pattern design (d/a relations and cavity type) e-beam and develop; dryetch; undercut and resulting pillar; fab deviations from design. Tunneling Schottky barrier ohmic contact description. Ge-doped-Au deposition on n-region and Au on p-region works because Ge is n-dopant in III-V and after annealing n-contact is very good; Au-on-p-region automatically forms a good ohmic contact due to fermi level pinning [see Miller 243 notes p.129].

\section{Experiment}
The structures are mounted in a confocal microscope setup which allows for independent positioning of two laser beams: a narrow-linewidth continuous-wave (cw) probe beam at a  wavelength that is tunable from 1250-1369\unit{nm}; and a pump beam for exciting photoluminescence (PL) with a wavelength of $633$\unit{nm}. A movable pinhole in the image plane of the confocal microscope setup allows us to collect light from different regions of the chip with a size as small as $\sim 3\mu$m.

\subsection{Device Characterization}
\begin{figure}
\centering{\includegraphics[width=0.7\linewidth]{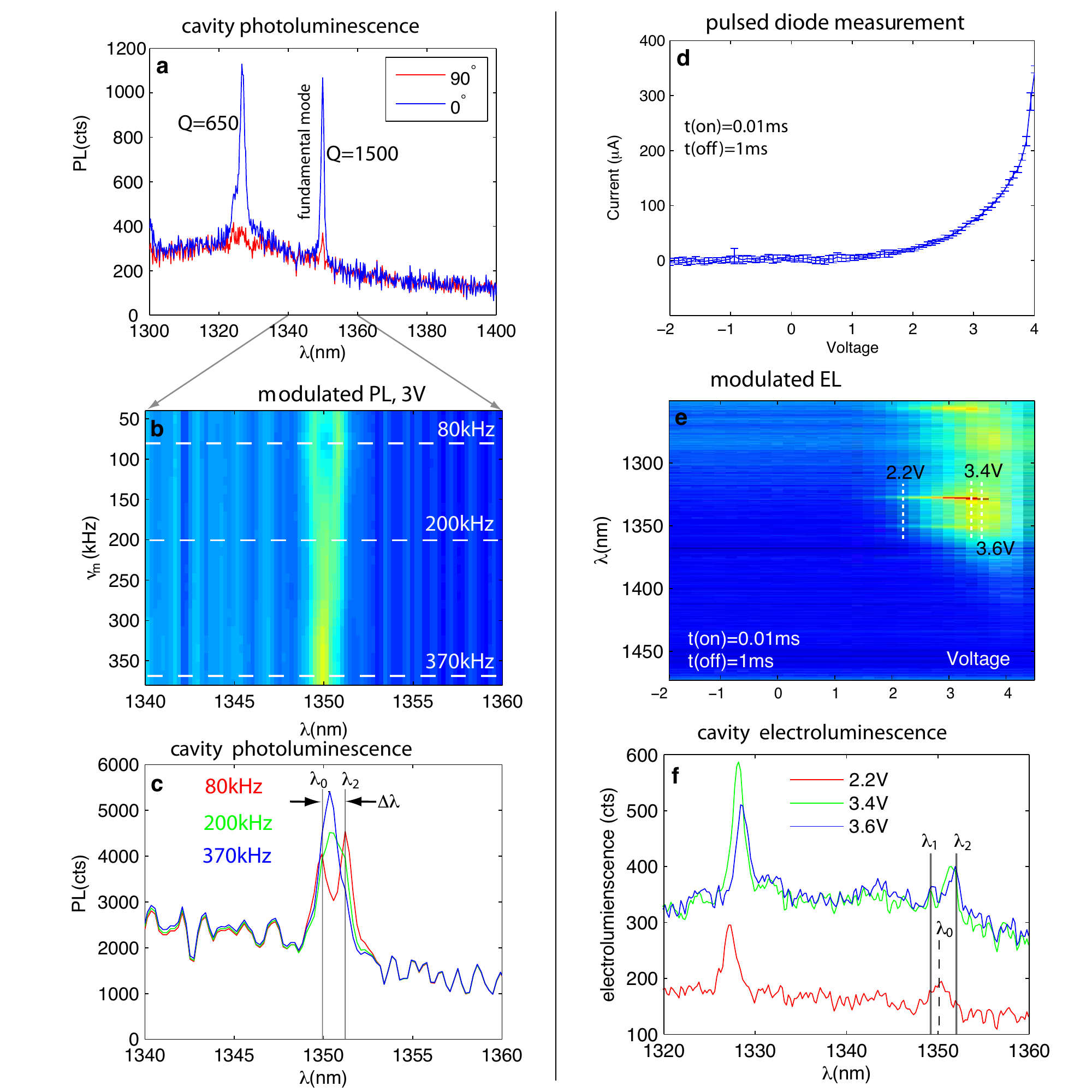}}
\caption{{\footnotesize (a) PL of cavity under above-band excitation with 633 nm laser. (b) PL collected from cavity as function of the frequency of electrical control pulses at a forward bias of 3V. (c) PL under electrical injection at 80kHz; 200kHz; and 370kHz. (d) Measured current under pulsed excitation when the above-band laser is off. (e) EL corresponding to (d), measured on a spectrometer. (f) The EL spectra corresponding to the dashed lines in (e) show that the cavity resonance splits into two discernable peaks at higher voltage. }} \label{fig:charact}
%\vskip2mm
\end{figure}

\begin{figure}
\centering{\includegraphics[width=.9\linewidth]{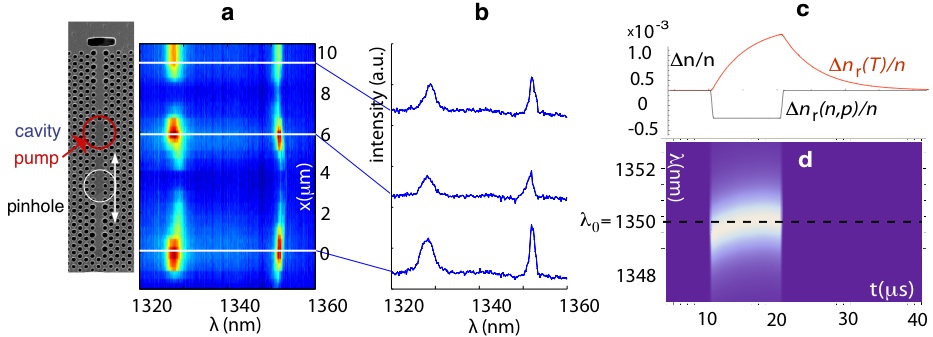}}
\caption{{\footnotesize (a) PL when the cavity is pumped by a focused laser and the collection pinhole is scanned across the length of the device. (b) The spectra at the top coupler; the cavity; and the bottom coupler.  The intensity through the bottom coupler is slightly larger because the second-order grating has four periods, whereas the top grating only has three. (c) When a $10\mu$s square-wave voltage with amplitude 3V is applied across the cavity, the cavity refractive index shifts by both free carrier injection and temperature change, which are estimated in this plot. (d) Expected electroluminescence when the cavity is pumped with the same $10 \mu$s long square-wave. The cavity first rapidly blue-shifts away from the cold-cavity resonance $\lambda_0$ due to free carrier injection, and then red-shifts over a longer time scale $\sim 5\mu$s due to heating.}} \label{fig:scan_PL}
%\vskip2mm
\end{figure}

We first characterize the photonic circuit by QD photoluminescence that is excited using the $633$\unit{nm} pump laser. Fig.\ref{fig:charact}(a) shows the cavity spectrum observed when the cavity is optically pumped and the observation pinhole is open to collect from the full structure. We observe two cavity modes at wavelengths $\lambda =1350$ nm $(Q=1500)$ and $\lambda' =1327$ $(Q'=650)$, both polarized perpendicular to the long cavity axis.   We identify these modes as the fundamental and first-order modes of the L3 cavity\cite{2007.APL.UK_ppl.mode_structure_L3}. We will from now on concentrate on the fundamental cavity mode at $\lambda \approx 1350$ nm. To characterize the transfer of the cavity emission to the waveguide/grating couplers, we graph in Fig.\ref{fig:scan_PL} the PL when the cavity is pumped while the pinhole is scanned across the length of the device. The QD PL that is collected directly from the cavity is lower than the PL collected at either end of the waveguides.  Assuming no material loss in the waveguides, we can use this measurement to estimate that the coupling efficiency out of the two gratings is about 40\%, based on a coupled mode equations model described in Appendix \ref{app:coupled_modes}. 

% Here I could compare the emission from teh cavity with pinhole closed to pinhole open. This will roughly indicate the fraction of recombination occurring in the cavity. (actually I can model this exactly with what I did for the APL perturbation paper). ...x..x.xxxx...xxx.... Hence the fraction is yy.  

We use the electroluminescence (EL) from the p-i-n diode to characterize the electrical pumping of the structure. Fig.\ref{fig:charact}(e) plots the EL when the pinhole is closed around the outcoupling grating. The voltage is pulsed at 1\unit{kHz} with a 1\% duty cycle to reduce heating. We observe both the fundamental and higher-order modes at a voltage above $2$\unit{V}, which corresponds to a current of 18 \unit{$\mu$A} (see Fig.\ref{fig:charact}(d)). From simulations, we estimate that the carriers distribute rather evenly across the whole membrane, as seen in Fig.\ref{fig:setup}(b). The cavity modes are visible above the background EL because the cavity-coupled QDs emit more rapidly through the Purcell effect\cite{2005.PRL.Englund,2005.Science.Noda.SE_control}. Above a voltage of $4$\unit{V}, the EL drops rapidly because of heating of the membrane. 

We will now use the EL spectra to estimate the cavity index change due to carriers, $\Delta n_r(n,p)$, and temperature, $\Delta n_r(T)$:\begin{equation}
\label{eq:dn}
\Delta n_r = \Delta n_r(n,p) +\Delta n_r(T), 
\end{equation}
where $n$ and $p$ are the electron and hole concentrations, respectively, and $T$ is the temperature. For the carrier-dependent term, we consider contributions due to bandgap narrowing ($ \Delta n_r(n,p)_{BG}$), bandgap filling ($ \Delta n_r(n,p)_{BF}$), and free carrier effects ($ \Delta n_r(n,p)_{FC}$). As derived in Appendix \ref{app:n}, the latter two contributions are dominant, and we approximate their combined effect on $n_r$ as 
\begin{eqnarray}\label{eq:Dnn}
 \Delta n_r(n,p) &\approx & \Delta n_r(n,p)_{BF} + \Delta n_r(n,p)_{FC} \\ \nonumber
% &=& -9.4 \cdot 10^{-21} (\Delta n \cdot \mbox{cm}^{3}) - 14.1 \cdot 10^{-21} (\Delta p \cdot \mbox{cm}^{3}) 
  &=& -5.4 \cdot 10^{-21} (\Delta n \cdot \mbox{cm}^{3}) - 2.5 \cdot 10^{-21} (\Delta p \cdot \mbox{cm}^{3}) 
 \end{eqnarray}
 The temperature-dependent index change is modeled as 
 \begin{equation}\label{eq:DnT}
 \Delta n_r(T)= 8.4\cdot 10^{-4} \Delta T,
 \end{equation}
 where $\Delta T$ is the temperature change from 300 K\cite{1995.APL.Talghader.thermal_nr_GaAs,2000.APL.Corte.nr_GaAs}. 

Experimentally, we can deduce the dielectric index change $\Delta n$ from the frequency shift $\Delta \omega_c$ in the cavity resonance, using second order perturbation theory \cite{Johnson2002} 
\begin{equation}\label{eq:dw_dn}
\Delta\omega_c \approx -\fr{\omega_c}{2} \fr{\int \Delta\epsilon \| \vec{E} \|^{2}}{{\int \epsilon \| \vec{E} \|^{2}}}
\end{equation}
From FDTD simulations, we note that the cavity field is primarily in the high-index material. We can then approximate, for a small index shift $\Delta n$, that 
\begin{equation}
\label{eq:dl_n}
\fr{\Delta \lambda_c}{\lambda_c}  \approx  \fr{\Delta n}{n_0}, 
\end{equation}
where $n_0 = 3.4$ is the index of GaAs at a wavelength of 1.3 $\mu$m \cite{Palik-Handbook}. Eqs. [\ref{eq:Dnn},\ref{eq:DnT}] show that temperature and carrier-mediated shifts on the cavity resonance are expected to be competing effects. 

Fig.\ref{fig:charact}(f) plots the EL spectra at the voltages indicated in Fig. \ref{fig:charact}(e).  As the voltage is increased from 2.2V to above 3V, the cavity resonance separates from $\lambda_0$ into two peaks centered at $\lambda_1,\lambda_2$, which are split by 2.2\unit{nm} (the splitting is not visible for the higher order mode at 1327\unit{nm} because of the lower $Q$).  $\lambda_1$ appears slightly blue-detuned from $\lambda_0$, which would indicate index modulation by free-carriers and/or band-filling. The carrier shift is expected to occur during the RC-limited response time $\tau_{RC}\sim 3$\unit{ns}, as described later. The tuning is accompanied by heating of the structure which causes the red-shift to $\lambda_2$. We estimate that the thermal effect occurs on the time scale of $\sim 5\mu$s (estimated from varied pulse-length measurements).  From the red shift, we calculate a temperature-induced refractive index shift of $ \Delta n_r(T) \approx 1.3\cdot 10^{-3}$ and a corresponding $\Delta T \approx  1.56 ^{\circ}$C. The blue-shift indicates $ \Delta n_r(n,p) \approx -1\cdot 10^{-3}$. 

We compare these experimentally obtained index variations to numerical simulations.  From the carrier simulations in Fig.\ref{fig:setup}(b), we estimate that at the location of the cavity and at $V_{in}=3$\unit{V}, $ \Delta n_r(n,p) \approx -2 \cdot 10^{-3}$, averaged over the membrane thickness, which is fairly close to our observation. The temperature simulation predicts $\Delta n_r(T)\approx 2.2\cdot 10^{-3}$, which is also reasonably close to our observation. A time-dependent simulation of the carrier and temperature index shifts gives the cavity evolution after the control pulse is turned on (Fig.\ref{fig:scan_PL}(d)). This simulation indicates that the cavity first rapidly shifts to short wavelengths due to the carrier-induced refractive index change, and then shifts to longer wavelength because of heating. 

We will now consider how fast the cavity can be shifted. Figs.\ref{fig:charact}(b,c) plot the photoluminescence when the cavity is excited with the 633\unit{nm} pump laser and the structure is electrically modulated with a 0-3 V square wave at a modulation rate $\nu_m$ and duty cycle of $20\%$. The integration time is $100$\unit{ms} --- much longer than the switching time. At low frequency, we observe a cavity splitting of $\Delta\lambda_c \approx 1.30$\unit{nm}, which indicates a refractive index change of the cavity $\Delta n/n\approx 9.6\cdot 10^{-4}$. The splitting blurs at a driving frequency of $\nu_{m}\sim 150-300 $\unit{kHz}; we speculate that this occurs because the cavity does not reach steady-state temperature during each pump period. Above 300 kHz, the blurred feature narrows into a single peak that is red-shifted by a constant $\Delta \lambda = 0.4$\unit{nm} from $\lambda_0$. At this range of modulation frequencies, the temperature fluctuations decrease as the modulation is faster than the thermal response time, and the cavity remains at a constant temperature-induced offset. Although we expect there to be a shift in the resonance due to free carrier and/or band-filling effects, this shift is considerably smaller than the cavity linewidth and could not be verified even at the highest modulation amplitude of 4V. 

\subsection{Optical modulation through the photonic crystal cavity}
\begin{figure}
\centering{\includegraphics[width=.6\linewidth]{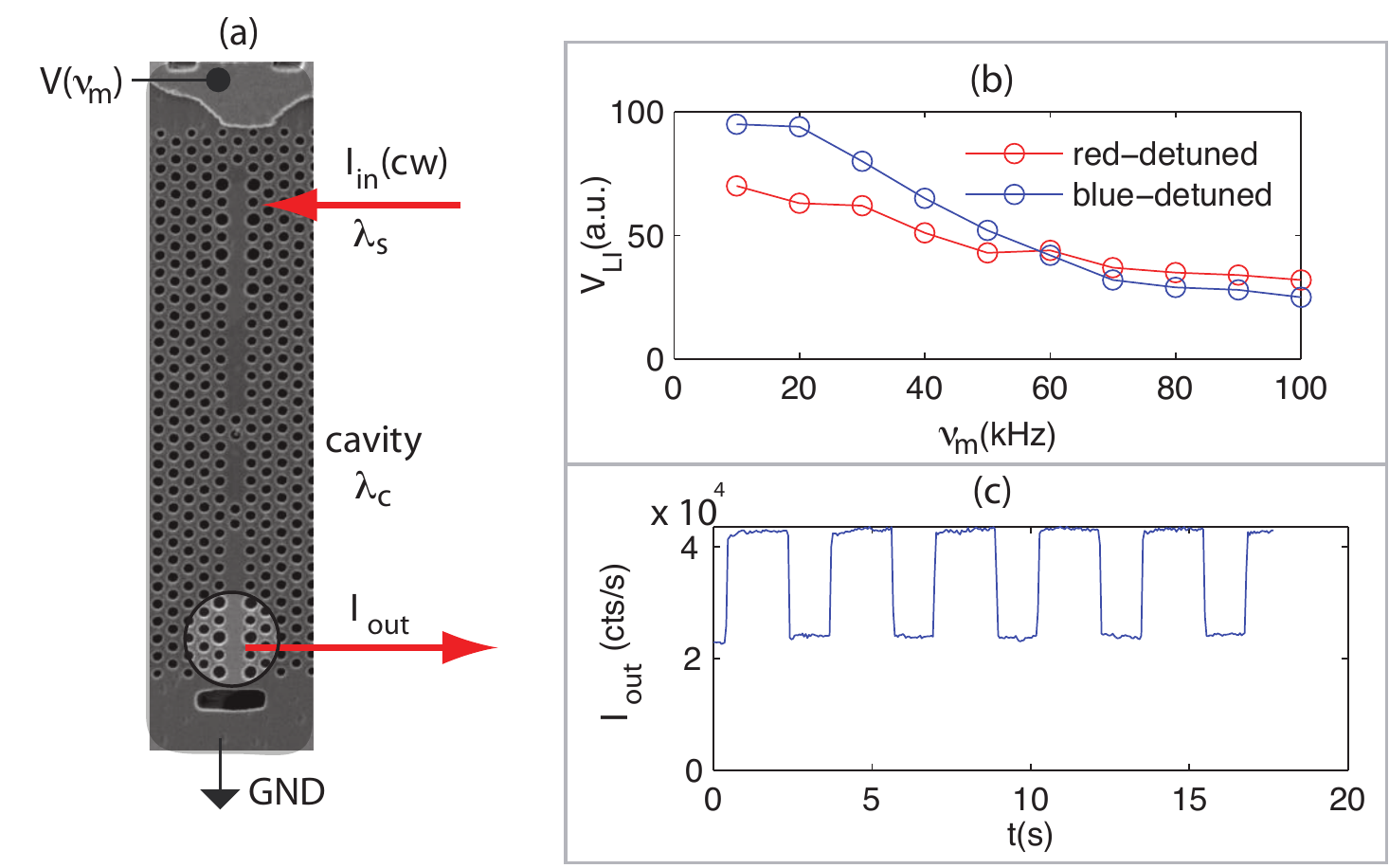}}
\caption{{\footnotesize (a) Setup for cavity transmission experiment. The laser at wavelength $\lambda_s$ is coupled into the input grating; through the cavity; and out-coupled through through the bottom grating, which is selected with a pinhole. (b) The signal is modulated at $v_m$ and measured using lock-in detection, giving voltage $V_{LI}$. We measured $V_{LI}$ for $\lambda_s$ red-detuned and blue-detuned from the zero-voltage cavity frequency. (c) Cavity transmission measured in time on a spectrometer. } }\label{fig:modulation_rate}
%\vskip2mm
\end{figure}

We will now operate the cavity as a switchable drop filter. As illustrated in Fig.\ref{fig:modulation_rate}(a), an external laser is coupled through the input grating into the waveguide.  If it is on resonance with the cavity frequency, it is transmitted to the output waveguide and scattered by the grating towards the objective lens.  To reduce stray light, we observe the transmission in the crossed polarization: the input beam is polarized at $45^{\circ}$ to the waveguide, while the output is observed at $-45^{\circ}$. The output is also spatially filtered using the pinhole to select the output grating. Fig.\ref{fig:modulation_rate}(c) shows the transmission observed on the spectrometer when the signal beam's wavelength $\lambda_s= 1351$ nm, which is red-detuned $1$\unit{nm} from the cavity frequency at zero bias field.  We also measured the transmission when the laser was blue-detuned by $\sim 0.5$\unit{nm} from the zero-bias cavity wavelength. Because the transmitted signal intensity was small due to low coupling through the input grating, it was necessary to measure the transmitted amplitude using a lock-in amplifier. This limited the measurement to the lock-in amplifier's cut-off frequency of 100\unit{kHz}. Fig.\ref{fig:modulation_rate}(b) shows the modulation amplitude observed on the lock-in amplifier. In addition to the lock-in amplifier restriction, the modulation rate is limited by the frequency-dependent thermal effects of the device, as discussed above. The thermal stability could be greatly improved in future experiments by placing the cavities on top of a low-index substrate such as sapphire or silicon dioxide for improved thermal conductivity\cite{2000.APL.Hwang,2001.ElectronLett.Monat,2007.OpEx.Raj,2006.OpEx.Bakir-Fedeli}, or by using line coding schemes \cite{2002.OSA.Bass.FiberOptics}. The thermal stabilization in the PL under modulation exceeding 200 kHz suggests operating beyond this frequency with a faster detection technique. To estimate the ultimate modulation speed, we measured the RC time constant.  The capacitance was directly measured to be only 3 pF, while the resistance was estimated at 1.1 k$\Omega$ from the forward bias part of the I-V curve. The RC time constant of $\sim 3 $ns suggests that a fast modulation speed is possible.

%We will now consider if absorption of the incident beam by free carriers is signficant.  The free carrier absorption inside the waveguide and cavity is XXX. Based on the free carrier modulation, we expect an absorptive modulation of XXX. This is much smaller than the observed modulation. Thus, the incident beam is reflected from the cavity rather than absorbed. An advantage of the reflective modulator is that it can handle high optical power. Additionally, the switchable cavity could be used to route signals between two output ports (cite?). 

%Next paragraph: Modulation: heat and thermal. 
%A. Demonstrate modulation
%Equipment: laser, spectrometer, oscilloscope, high speed NIR detector
%-eye diagrams: oscilloscope at highest speed/any voltage that works
%-speed - highest speed that opens eye at any voltage -- DON"T HAVE THIS.
%-voltage: bias and swing required to open eye, quote extremes (lowest operating voltage and voltage for best eye diagram and speed
%B. characterize modulation: use absorption measurements + Kramers Kronig to determine change in refractive index
\section{Conclusions}
In conclusion, we have demonstrated a modulator that relies on a photonic crystal nanocavity as the active component. The small size allows for a low capacitance and promises operation at high bandwidth. While we measured $1/RC\sim 300$MHz, the modulation speed in future designs could be increased significantly by changing the refractive index only in the photonic crystal cavity whose area is less than 100 times smaller than the full membrane in this study. Lateral dopant implantation could allow a small junction with sub-fF capacitance and a time constant of $RC<10$ ps\cite{2006.APL.Schmidt}.  The frequency-selective modulation of the cavity is suited for wavelength division multiplexing, which greatly increases the total interconnect bandwidth and may become essential in off-chip optical interconnects\cite{2009.Miller}.  Electrically controlled photonic crystal networks furthermore have promise in applications including biochemical sensing \cite{2004.OptLett.Girolami,2003.APL.Loncar-Scherer.biosensor_PC} and  quantum information processing in on-chip photonic networks \cite{2007.OpEx.Englund,2008.Science.OBrian.quantum_circuit}. 

% ADD: deflective switch.. 

\subsection{Acknowledgements}
This work was supported by the MARCO Interconnect Focus Center and the and DARPA Young Faculty Award. 

\appendix
\section{Sample growth}\label{app:growth} 
The sample is grown by molecular beam epitaxy on an $n$-type GaAs substrate and consists of a 1\unit{$\mu$m} $n$-doped Al$_{0.8}$Ga$_{0.2}$As sacrificial layer, a $40$\unit{nm} $n$-doped GaAs layer, a 160\unit{nm} intrinsic GaAs membrane that contains three layers of InAs quantum dots (QDs) separated by 50\unit{nm} GaAs spacers, a 25\unit{nm} $p$-doped GaAs layer and a 15\unit{nm} highly $p$-doped GaAs layer to ensure low-resistance contacts. The QD layers were grown by depositing 2.8 monolayers (ML) of InAs at 510$^{\circ}$C at a growth rate of 0.05 ML/s. To achieve emission at $\sim1.3$\unit{$\mu$m}, the dots were capped with a 6\unit{nm} In$_{0.15}$Ga$_{0.85}$As strain-relaxing layer.

\section{Fabrication}\label{app:fab} 
First, metal contacts are deposited on the n-type substrate and annealed to form the bottom contacts. These contacts consist of Au/Ge/Ni/Au layers. Then the photonic crystal structure, isolating layer, trenches, and the top-contact PMMA mask are fabricated by a combination of electron beam lithography and dry/wet etching steps, which are outlined in Ref.\cite{2005.PRL.Englund}. The p-contact consists of Pt/Ti/Pt/Au layers. The free-standing membrane is created by removing the sacrificial AlGaAs layer using a hydrofluoric acid-based selective wet etch. 

\section{Index dependence on carriers and temperature}\label{app:n}
The primary sources of refractive index change in the GaAs membrane are bandgap narrowing, bandgap filling, free carrier effects, and temperature effects. We neglect band-gap narrowing as it only changes slowly with carrier density at high carrier densities and is far weaker than band-filling effects at 1350 nm\cite{1990.JQE.Bennett.carrier_induced_nr_GaAs}. Because the carrier effects are independent, we can express $\Delta n_{r}$ as the sum due to band filling, free carrier effects, and thermal changes, respectively: 

% the bandgap narrowing contribution should be positive and equal to ~ 5e-3 for n=10^18 --- check impact on total Dn and comparison to the model! 

\begin{equation}
\Delta n(n,p,T)=\Delta n(n,p)_{BF}+\Delta n(n,p)_{FC}+\Delta n(T)
\label{eq:1}
\end{equation}\
We follow the discussion in Ref.\cite{1990.JQE.Bennett.carrier_induced_nr_GaAs} to estimate the carrier contributions $\Delta n(n,p)_{BF}+\Delta n(n,p)_{FC}$. Using fundamental constants\cite{1990.JQE.Bennett.carrier_induced_nr_GaAs}, a DC refractive index  $n_{g}=3.6$, normalized electron and hole masses $m_{e}=0.066$, $m_{h} =0.45$, and assuming equal $\Delta n$ and $\Delta p$, 
%\begin{equation}
%\Delta n_{r}(n,p)_{FC}=-\frac{ 6.9\times10^{-22}}{n_{g}E^{2}} \left(\frac{\Delta n}{m_{e}}+\frac{\Delta p}{m_{h}}\right)
%\end{equation}
\begin{eqnarray}
\Delta n_{r}(n,p)_{BF}& \approx&-2\times10^{-21} \left(\Delta n + \Delta p\right) \\
\Delta n_{r}(n,p)_{FC}&=&-\frac{ e^2\lambda^2}{8\pi^2c^2\epsilon_{0}n_{0}} \left(\frac{\Delta n}{m_{e}}+\frac{\Delta p}{m_{h}}\right)\\
&\simeq&-3.4\times10^{-21}\Delta n  -5.0\times10^{-22}\Delta p
\label{eq:5}
\end{eqnarray}
We note also that larger refractive index changes are possible if working with photon energies nearer to the band edge \cite{1981.PRL.Miller.bandgap_resonant_NL} or by exploiting the index change in quantum confined structures \cite{1988.IEEE.Miller.FE_QW_absorption}.  The temperature dependence of the refractive index follows
\begin{equation}
\Delta n_{r}(T)= n_{0}\cdot \alpha(T-T_{0}),
\end{equation}

where $T_{0}$=300K is the initial temperature and $T$ is the operating temperature.

We chose $n_{r}$=3.410 for GaAs at $\lambda=1.3{\mu}$m T=300${^\circ}$C, after \cite{Palik-Handbook}.  $\alpha$ is the thermo-optic constant, which we take to be  $2.5\times10^{-4}/^{\circ}$C as an interpolation between Refs.\cite{1995.APL.Talghader.thermal_nr_GaAs,2000.APL.Corte.nr_GaAs} for $\lambda=1.35{\mu}$m.

Therefore, the total expected change in refractive index is

\begin{equation}
\Delta n_{r}(n,p,T)=-5.4\times10^{-21}\Delta n-2.5\times10^{-21}\Delta p+8.5\times10^{-4}\Delta T
\end{equation}

%\begin{table}[h]
%\caption{GaAs thermo-optic constant}
%\begin{center}
%\begin{tabular}{| c | p{2.5cm} | p{2.5cm}  | p{2.5cm} |     }

%\hline 
% & Talghader${^1}$ &  Corte ${^2}$  & Our simulation \\ 
%\hline

% $(dn/dT)_{GaAs}$ 
% & $(2.67\pm0.07)\times10^{-4}/^{\circ}$C 
% & $(2.35\pm0.07)\times10^{-4}/^{\circ}$C 
% & $2.5\times10^{-4}/^{\circ}$C
%\\ \hline
%Wavelength & 960-1030nm & 1500nm & 1300nm
%\\ \hline
%Temperature & 25-150$^{\circ}$C & 27-327$^{\circ}$C & 300$^{\circ}$C
%\\ \hline
%\end{tabular}
%\end{center}
%\end{table} 

%\subsection{Free carrier absorption}
%The free carrier absorption in the waveguide and cavity are estimated by the total time spent in both. For the waveguide, this is $t_w=L_w/v_g$, where $v_g$ is the group velocity near the $k_x=\pi/a$ point; for the cavity, the time is $t_c=Q/\omega\sim 1.1$\unit{ps}.  Assuming $t_w\leq t_c$, the transmission is $T\geq e^{-\alpha (2 t_c)(c/n_r)} \approx xxx$, where $\alpha=xxx$ is the free carrier absorption.  When $V_{in}=0$, the average concentration of electron and hole carriers of $\sim 10^{18} $/cm$^{3}$ gives  $\alpha \sim XXX$ and corresponding $T(0\mbox{V})=XXX$. When $V_{in}=2$\unit{V}, the average concentration of electron and hole carriers of $\sim 10^{18} $/cm$^{3}$ gives  $\alpha \sim XXX$ and corresponding $T(2\mbox{V})=XXX$.

\section{Photonic circuit model and characterization}\label{app:coupled_modes}
Let $a,b,c,d,e$ represent, respectively, the input grating; first waveguide; cavity; second waveguide; and output grating. With $\kappa_g,\kappa_{w,g},\kappa_{w,c},\kappa$ representing, respectively, the vertical coupling rate through the grating; the cavity-grating coupling rate; the cavity-waveguide coupling rate, and the cavity loss rate, we have 
\begin{eqnarray}
\D{a}{t} & = & -\kappa_g a - i \kappa_{w,g} b \\ \nonumber
\D{b}{t} & = & -i\kappa_{w,g} a - i \kappa_{w,c} c \\  \nonumber
\D{c}{t} & = & -\kappa c - i \kappa_{w,c} b - i \kappa_{w,c}d + p(t) \\ \nonumber
\D{d}{t} & = & -i\kappa_{w,g} e - i \kappa_{w,c} c \\ \nonumber
\D{e}{t} & = & -\kappa_g e - i \kappa_{w,g} d, \\  \nonumber
\end{eqnarray}
where $p(t)$ is the pump rate. We solve this system of equations in the steady state, since the coupling rates are much faster than changes in $p(t)$. Then we can solve for the transfer efficiency to the two grating couplers as $\eta_c=2 (\kappa_g a)^{2}/(2(\kappa_g a)^{2}+(\kappa c)^{2})=(1+(\kappa \kappa_{w,g}/\sqrt{2}\kappa_g\kappa_{w,c})^{2})^{-1}$.

Assuming negligible loss the waveguide, we can estimate from the measured intensities in Fig.\ref{fig:scan_PL}(b) that $\eta_c\approx 0.4$ and $\kappa \kappa_{w,g}/\kappa_g \kappa_{w,c}=1.8$. Assuming $\kappa/\kappa_{w,c}\sim 0.85$ from simulation, we estimate $\kappa_g/\kappa_{w,g}\sim 0.5$. An increase in $\kappa_{w,g}$, which could be achieved by fabricating larger grating couplers, is therefore expected to improve the coupling efficiency to the input/output ports.

\end{document}